\documentclass[conference,11pt]{IEEEtran}
	\IEEEoverridecommandlockouts
	\usepackage{cite}
	 \usepackage{amsmath,amssymb,amsthm,amsfonts,multicol, nccmath}
	\usepackage{mathtools}
	\usepackage[noend]{algpseudocode}
	\usepackage{amssymb,epsfig,color,cite,epstopdf}
	\usepackage{amsmath,amsthm,thmtools}
	\usepackage{amssymb}
	\usepackage{graphicx}
	\usepackage[tight]{subfigure}
	\usepackage{caption}  
	\usepackage{stackengine}
	\usepackage{stfloats}

	\usepackage[linesnumbered,ruled]{algorithm2e}
	\usepackage{tabularx,tabulary,multirow,adjustbox,booktabs}
	\usepackage[flushleft]{threeparttable}
	\usepackage{array,bigdelim,blkarray}
	\usepackage{multicol,multirow}
	\usepackage{url}
	\usepackage{bm}
	\usepackage{makecell,pict2e}
	\usepackage{diagbox}
	\usepackage{stmaryrd}
	\usepackage{framed}
	\usepackage{mdframed}
	\usepackage{tcolorbox}
	\usepackage{slashbox}
	\usepackage{diagbox}
	\usepackage{balance}
	\usepackage{pifont}
	\usepackage{mathdots}
	\usepackage{MnSymbol}

	\usepackage{tikz}
	\usepackage{pgfplots}
	\usetikzlibrary{shapes}

	\newcommand{\hwplotD}{\raisebox{2pt}{\tikz{\draw[-,green,solid,line width = 2pt](0.,0.07mm) -- (3.5mm,0.07mm)}}}

\def\x{{\mathbf x}}

\def\z{{\mathbf z}}
\def\e{{\mathbf e}}

\def\w{{\mathbf w}}
\def\X{{\mathbf X}}
\def\s{{\mathbf s}}
\def\a{{\mathbf a}}

\def\A{{\mathbf A}}

\def\C{{\mathbf C}}

\def\R{{\mathbf R}}
\def\I{{\mathbf I}}

\def\U{{\mathbf U}}

\def\S{{\mathbf S}}
\def\X{{\mathbf X}}

\def\E{{\mathbf E}}

\def\0{{\mathbf 0}}

\newcommand{\RR}{\mathbb{R}}

\newcommand{\TT}{\mathcal{T}}

\newcommand{\diag}{\operatorname{\diag}}

\newcommand{\argmin}{\mathop{\rm argmin}}
\newcommand{\argmax}{\mathop{\rm argmax}}

\def\dashdotted{\xleaders\hbox to 1em{$- \cdot$} \hfill $-$}

\newcommand*{\Scale}[2][4]{\scalebox{#1}{$#2$}}%

\makeatletter
\newcolumntype{M}[1]{>{\hbox to #1\bgroup\hss$}l<{$\egroup}}

\newcommand\@brcolwidth{3em}

\def\@brarray[#1]{\array{r*\c@MaxMatrixCols {M{#1}}}}
\makeatother

\begin{document}
		
	\title{Robust Sparse Subspace Tracking from  Corrupted Data Observations \\[-0.25em]}
	

	\author{\IEEEauthorblockN{Ta Giang Thuy Loan\IEEEauthorrefmark{1}\IEEEauthorrefmark{5}, Hoang-Lan Nguyen\IEEEauthorrefmark{1}\IEEEauthorrefmark{5}, Nguyen Thi Ngoc Lan\IEEEauthorrefmark{1}, Do Hai Son\IEEEauthorrefmark{2}, \\ Tran Thi Thuy Quynh\IEEEauthorrefmark{1},  Nguyen Linh Trung\IEEEauthorrefmark{1}, Karim Abed-Meraim\IEEEauthorrefmark{4},  Thanh  Trung Le\IEEEauthorrefmark{1}\IEEEauthorrefmark{5}  \\ \\}
	    \IEEEauthorrefmark{1} VNU University of Engineering and Technology, Hanoi, Vietnam \\
	    \IEEEauthorrefmark{2} VNU Information Technology Institute, Hanoi, Vietnam \\
	    \IEEEauthorrefmark{4} PRISME Laboratory, University of Orleans, Orleans, France \\   
	    \thanks{\IEEEauthorrefmark{5} Thuy Loan and Hoang Lan are co-first authors.  Corresponding author: Thanh Trung Le (thanhletrung@vnu.edu.vn). This work was funded by the Postdoctoral Scholarship Programme of Vingroup Innovation Foundation (VINIF), code VINIF.2024.STS.40. }
	    \vspace{-2em}
	}
	
	\maketitle
	
	\begin{abstract}
	Subspace tracking is a fundamental problem in signal processing, where the goal is to estimate and track the underlying subspace that spans a sequence of data streams over time. In high-dimensional settings, data samples are often corrupted by non-Gaussian noises and may exhibit sparsity. This paper explores the alpha divergence for sparse subspace estimation and tracking, offering robustness to data corruption. The proposed method outperforms the state-of-the-art robust subspace tracking methods while achieving a low computational complexity and memory storage. Several experiments are conducted to demonstrate its effectiveness in robust subspace tracking and direction-of-arrival (DOA) estimation.  
	 
	\end{abstract}
	
	
	\vspace{-0.5em}
	 \section{Introduction}
	 	\vspace{-0.25em}
	\label{sec:intro}

	\setlength{\abovedisplayskip}{3pt}  
	\setlength{\belowdisplayskip}{3pt} 
	\setlength{\jot}{0.25pt}

	With the ubiquity of big data streams, modern online applications are continuously generating massive volumes of high-velocity data~\cite{kolajo2019big}. The dynamic and evolving nature of such data streams presents significant challenges for conventional data mining techniques, which often assume access to static or batch data.  In many scenarios, data samples arrive sequentially, and it is desirable to update subspace estimates in (near) real time without revisiting old observations. Subspace tracking (ST) addresses this need by estimating and continuously updating a low-dimensional subspace that captures the underlying structure of streaming data~\cite{delmas2010subspace}. 
	
	Modern data are often incomplete, unreliable, or corrupted due to collection processes, inconsistencies, the presence of non-Gaussian noise, and sparse outliers~\cite{fan2014challenges}. 
	High dimensionality further exacerbates these issues by increasing computational and memory demands, while also degrading algorithmic performance. A principled and effective way to mitigate these challenges is to exploit the empirical observation that high-dimensional data streams often lie near a low-dimensional and sparse subspace that can evolve over time~\cite{8398586,dung2021robust}. In the presense of data corruption, the classical subspace tracking problem naturally extends to the more general and challenging task of robust sparse subspace tracking.
	
	Several ST algorithms have been proposed in recent years to address the challenges posed by data corruption~\cite{dung2021robust}. Broadly, these methods fall into two categories: (ii) algorithms designed to handle outliers and missing data, and (ii) algorithms designed to cope with abrupt or impulsive noise. The first category includes several notable approaches such as Grassmannian-based methods (e.g., GROUSE~\cite{balzano2010online}, GRASTA~\cite{he2012incremental}, pROST~\cite{seidel2014prost}), recursive least squares-based methods (e.g., PETRELS~\cite{6605610}, PETRELS-ADMM~\cite{9381678}, PETRELS-CFAR~\cite{dung2018}), recursive projected compressive sensing-based methods (e.g.,  ReProCS's variants \cite{zhan2016online,narayanamurthy2018provable,narayanamurthy2020fast}), and adaptive projected subgradient methods (e.g.,~\cite{6854654, chouvardas2015robust, slavakis2013adaptive}). Although these approaches are effective in handling outliers and missing entries, they typically rely on the assumption of slowly varying subspaces over time, making them less robust to abrupt changes or impulsive noise. 
	
	The second group includes robust variants of the Projection Approximation Subspace Tracking (PAST) algorithm (RPAST~\cite{chan2006robust}, MCC-PAST~\cite{zhang2015robust}, TRPAST~\cite{rekavandi2023trpast}); robust variants of adaptive or online power iteration methods (e.g., $\alpha$FAPI~\cite{thanh2023aFAPI},  ROBUSTQR~\cite{nguyen2022robust}); adaptive Kalman filtering (KFVM~\cite{liao2010new}), weighted recursive least-squares method (e.g., ROBUSTA \cite{dung2018}). These methods are designed to deal with impulsive noise but are generally sensitive to sparse corruptions and missing data. It is also worth noting that the aforementioned methods are not inherently designed for sparse subspace tracking. In the literature, only few sparse ST methods have been proposed. Notable examples include OIST~\cite{wang2016online}, OVBSL~\cite{giampouras2017online}, and OPIT~\cite{le2024opit}. However, these methods are not specifically designed to handle data corruption, particularly non-Gaussian noise.
	This limitation motivates the development of a new sparse subspace tracking algorithm that can both exploit sparsity in the subspace structure and maintain robustness against various types of data corruption.
		\begin{table}[!t]
		\small  
		\centering
		\vspace{2em}
		\caption{Conventional Notations}\vspace{-0.5em}
		\label{tab:notations}
		\begin{tabular}{|l p{5.83cm}|}
			\hline
			& \\[-.5em]
			$x, \bm{x}, \bm{X}, \mathbb{X}$ & scalar, vector, matrix, and set/support \\
			$x_i$ or $\bm{x}(i)$ & $i$-th entry of $\bm{x}$ \\
			$x_{ij}$ or $\bm{X}(i,j)$ & $(i,j)$-th entry of $\bm{X}$ \\
			$\X(i,:)$, $\X(:,j)$ & $i$-th row and $j$-th column of $\bm{X}$ \\
			$\bm{X}^{\top}, \bm{X}^{-1}$ & transpose, inverse of $\bm{X}$ \\
			$\operatorname{QR}(\X)$ & QR decomposition of $\X$ \\
			$ \|\cdot\|_F$ & Frobenius norm \\
			$ \operatorname{tr}\{\cdot\}$ & trace operator \\
			$ \mathbb{E}\{\cdot\}$ & expectation  operator \\
			
			$\mathcal{N}(\mu, \sigma^2)$ & Gaussian distribution with mean $\mu$ and variance $\sigma^2$ \\
			$\mathcal{D}_\alpha\{g \| f\}$ & $\alpha$-divergence between two distributions $g$ and~$f$ \\[0.5em]
			\hline
		\end{tabular}
		\vspace{-0em}
	\end{table}
	In this paper, we introduce $\alpha$OPIT, an robust version of  OPIT that leverages an $\alpha$-divergence based weighting scheme for sparse subspace tracking with data corruption. Specifically, our method adaptively downweights anomalous/corrupted observations using $\alpha$-divergence, thereby enhancing its robustness against outliers and non-Gaussian noise. Additionally, $\alpha$OPIT incorporates a recursive covariance and hence subspace update with a forgetting factor, allowing for efficient online adaptation to evolving subspaces in streaming data. Experimental results indicate that $\alpha$OPIT consistently outperforms existing subspace tracking algorithms, particularly under challenging conditions involving non-Gaussian and impulsive noise, offering  robustness and improved tracking accuracy. 
	
	\emph{Paper Organization:} The rest of this paper is organized as follows. Section~II provides background on subspace tracking, the OPIT algorithm, and  $\alpha$-divergence. Section III introduces our  proposed method. Section IV presents the experimental results, and Section V concludes the paper. 
	For easy reference, Table~I summarizes frequently used notations in this paper.
	
	\vspace{-0.25em}
	\section{Background}
	\label{sec:background}
	\vspace{-0.25em}
		
	In this section, we begin by formulating the problem of robust subspace tracking, followed by a brief overview of the classical OPIT method. We then introduce the $\alpha$-divergence, which is employed to develop a weighting scheme that mitigates the influence of corrupted data.

	\vspace{-0.25em}
	\subsection{Subspace Tracking}
	\vspace{-0.25em}
	
	Assume that at each time $t$, we collect a data sample $\x_t \in \mathbb{R}^n$ which is generated under the following model
	\begin{align}\label{eq:data-model}
	\x_t = \A \w_t  + \bm \nu_t, \quad t = 1,2,\dots, T.
	\end{align}
	Here, $\A \in \mathbb{R}^{n \times r}$ is the underlying subspace matrix with $r < n$, $\w_t \in \mathbb{R}^{r}$ is  a weight vector, and $\bm \ell_t = \A \w_t$ represents the low-rank component of $\x_t$. The vector $\bm \nu_t \in \mathbb{R}^n$ denotes data corruption present in the observation. The problem of subspace tracking can be stated as follows:
	\begin{mdframed}
		\textbf{Subspace Tracking Problem}: On the arrival of a  new data sample $\x_t$ at each time~$t$, our goal is to estimate the underlying subspace $\A$ that spans the low-rank components $\{\bm \ell_i\}_{i=1}^t $.
	\end{mdframed}
	When the subspace matrix $\A$ is sparse, robust ST becomes robust sparse subspace tracking (SST) problem.\footnote{In nonstationary environments, the subspace matrix $\A$ can be slowly varying with time, i.e., $\A = \A_t$. Our method not only estimates it accurately but also effectively tracks its variation over time. See Fig.~2 for an illustration.}

	\vspace{-0.5em}
	\subsection{OPIT Method}
	\vspace{-0.25em}
	
	Online Power Iteration by Thresholding (OPIT) offers a fast method for tracking the underlying sparse subspace of data streams over time~\cite{le2024opit}. It builds upon the standard Power Iteration (PI) method for computing the dominant eigenvectors of the covariance matrix $\C_t = \mathbb{E}\{ \x_t \x_t^\top \}$. Particularly at the $k$-{th} iteration, PI updates 
	\begin{align} \label{eq:PI}
	    \S_\ell \leftarrow \C_t \U_{\ell-1}, \quad 
	    \U_\ell \overset{\text{Q-factor}}{\leftarrow} \operatorname{QR}(\S_\ell),
	\end{align} where $\operatorname{QR}(\cdot) $ denotes the QR factorization~\cite{delmas2010subspace}. PI starts from an initial matrix $ \U_0 \in \mathbb{R}^{n \times r} $ and returns an orthonormal matrix $\U_\ell$, where $L$ is the number of iterations. OPIT modifies PI by recurisvely updating the ``scaled" version of $\C_t$ and introducing a forgetting factor $ 0  < \beta \leq 1$ to exponentially discount the effect of old observations 
	\begin{align}\label{eq:Rt}
	\R_t = \beta \R_{t-1} +  \x_t \x_t^\top,
	\end{align} where $\C_t = t^{-1} \R_t$. According, OPIT rewrites the first step of~\eqref{eq:PI} as follows
	\begin{align} \label{eq:OPIT-St}
	    \S_t = \R_t\U_{t-1} = \beta \R_{t-1} \U_{t-1} +  \x_t \z_t^\top ,
	\end{align}
	where $\mathbf{z}_t = \U_{t-1}^\top  \x_t$. As small perturbations do not significantly affect the performance of power methods~\cite[Proposition 2]{abed2022sparse}, OPIT derives the following rule 
	\begin{align} \label{eq:OPIT_St_new}
	    \S_t \simeq \beta \S_{t-1} \E_{t-1} + \x_t \z_t^\top,
	\end{align} 
	where $\E_{t-1} = \U_{t-1}^\top \U_{t-2}$. After that, OPIT employs the threholding operation on~\eqref{eq:OPIT_St_new} before employing the QR factorization of $\S_t$, see Algorithm~\ref{al-thresholding}. We refer the readers to our work~\cite{le2024opit} for further details. Since OPIT is not inherently designed to handle data corruption, we incorporate the $\alpha$-divergence (introduced in the following subsection) to enhance its robustness.
	
	 {\LinesNumberedHidden
	 	\begin{algorithm}[!t]
	 	\footnotesize	
	 		\caption{\scshape \footnotesize Thresholding -  $\hat{\S} = \tau(\S,k)$}  
	 		\label{al-thresholding}
	 		\KwIn
	 		{Matrix $\S$ and a thresholding factor $k$}

	 		
	 		\SetKwBlock{Begin}{Main Procedure:}{End Main}
	 		\SetKwBlock{BeginIni}{Initialization: }{}
	 		\SetKwBlock{BeginThre}{Thresholding $\hat{\S} = \tau(\S_t,k)$:}{End}
		
	 		\Begin{ 
	 			$[n,r] = \operatorname{size}(\S)$\\
	 			\For{$i=1,2, \dots,r$}{
	 				
	 				$\s_i = \S(:,i)$ \\
	 				Select the set $\TT_t$ that  contains indices  of $k$ strongest entries (w.r.t. {absolute value})  of $\s_i$\\[-0em]
	 				Form $ \hat{\S}(j,i) = 
	 				\begin{cases}
	 					\s_i(j) &  \text{if  }  j \in \TT_t
	 					\\
	 					0 &  \text{if  }  j \not\in \TT_t
	 				\end{cases}$
	 			}
	 		}	
	 	\end{algorithm}	 
	 
	 }


	\vspace{-0.5em}
	\subsection{Alpha Divergence} 
	\vspace{-0.5em}
	
	The $\alpha$-divergence is a family of measures used to quantify the difference between two probability distributions~\cite{cichocki2010families}. Particularly given two distributions $g(\theta)$ and $f(\theta)$, its $\alpha$-divergence $D_{\alpha}\{g \,\|\, f\}$ is defined as 
	\begin{align} \label{eq:alpha}
	D_{\alpha} \big\{g \,\|\, f \big\} = \frac{1}{\alpha(1 - \alpha)} \left[ \int g(\theta)^{\alpha} f(\theta)^{1 - \alpha} \, d\theta - 1 \right],
	\end{align}
	where  $ 0  < \alpha < 1$   is a  tunable parameter to control the asymmetry of the divergence.\footnote{In the literature, there exist some other forms of $\alpha$-divergence, see~\cite{cichocki2010families} for more details.}
	Here, \eqref{eq:alpha} can be  considered as  a generalization of the well-known Kullback–Leibler (KL) divergence. In particular,  
	$ \lim_{\alpha \to 1} D_{\alpha}(g \,\|\, f) = \mathrm{KL}(g \,\|\, f)$ and 
	$\lim_{\alpha \to 0} D_{\alpha}(g \,\|\, f) = \mathrm{KL}(f \,\|\, g) $.
	The $\alpha$ divergence can offer robustness to several types of non-Gaussian noises which is exploited in the next section.

	\vspace{-0.5em}
	 \section{Proposed Method}
	\vspace{-0.5em}

	 {\LinesNumberedHidden
		\begin{algorithm}[!t]

			\caption{\scshape \footnotesize $\alpha$OPIT -   Online Power Iteration by Thresholding with $\alpha$ Divergence}  
	
	        \footnotesize
	        
			\label{al-aOPIT}
			\KwIn
			{$\{ \x_t \}_{t=1}^T, \x_t \in \RR^{n}$, rank $r$, a forgetting factor $0   \leq \lambda \leq 1$, alpha divergence with parameter $0 < \alpha < 1$, $0 < p \leq2$, and a thresholding factor $k${:}  \vspace{-0em}
				\begin{align} \notag
					~~~~k = \begin{dcases}
						\lfloor   (1-\omega_{{\mathrm{sparse}}}) n    \rceil & \text{if }  \omega_{{\mathrm{sparse}}} \text{ is given,} \\
						\lfloor   10 r \log n     \rceil & \text{if } \omega_{{\mathrm{sparse}}} \text{ is unknown,}
					\end{dcases}  \\[-1.5em] \notag
				\end{align} where $\omega_{{\mathrm{sparse}}}$ is the sparsity level of the subspace.}
			
	
			\SetKwBlock{Begin}{Main Procedure:}{End Main}
			\SetKwBlock{BeginIni}{Initialization: }{}
			\SetKwBlock{BeginThre}{Thresholding $\hat{\S} = \tau(\S_t,k)$:}{End}
			
			\BeginIni{
			 {Any $\U_0  \in \mathbb{R}^{n\times r}$}, $\S_{0}  = \bm 0_{n \times r}, $ and $ \E_0 = \I_{ r\times r} $
				
			}
			\Begin{			
							
				\For{$t =  1,2, \dots, T $}{
					$
					\begin{alignedat}{3}
						& \w_{t}  && = \U_{t-1}^\top \x_{t}  && \mathcal{O}(nr)   \\
						&\e_t && = \x_t - \U_{t-1} \w_t    && \mathcal{O}(nr)   \\
						& \omega_t && = \exp\Big(-{\frac{1-\alpha}{2} \| \e_t \|_F^p }\Big)  && \mathcal{O}(n) \\
				        & \S_{t}  && =  (1- \lambda) \S_{t-1} \E_{t-1} +  \lambda \omega_t \x_{t} \w_{t}^\top  \quad &&  \mathcal{O}(nr^2 + nr) \\
				        & \hat{\S}_{t}  && = \tau(\S_{t},k)\quad  &&  \mathcal{O}(nr + r k \log k) \\
				        & \U_{t} && = \operatorname{QR}(\hat{\S}_{t}) && \mathcal{O}(nr^2 ) \\
				        & \E_t &&= \U_{t-1}^\top \U_t && \mathcal{O}(nr^2 )
					\end{alignedat}
					$

				}
			}	
				
		\KwOut{$\U_T$}

		\end{algorithm}	 
	}

	In this section, we introduce a novel robust variant of OPIT,  called $\alpha$OPIT, which is designed to enhance OPIT's robustness and effectiveness against data corruption. The proposed method incorporates advanced weighting and subspace estimation techniques derived from $\alpha$-divergence and OPIT, allowing it to handle non-Gaussian noise and high-dimensional settings. In what follows, we detail how $\alpha$-divergence is incorporated into OPIT to improve its robustness.

\begin{figure}[!t]
		\centering
		\includegraphics[width=0.75\linewidth]{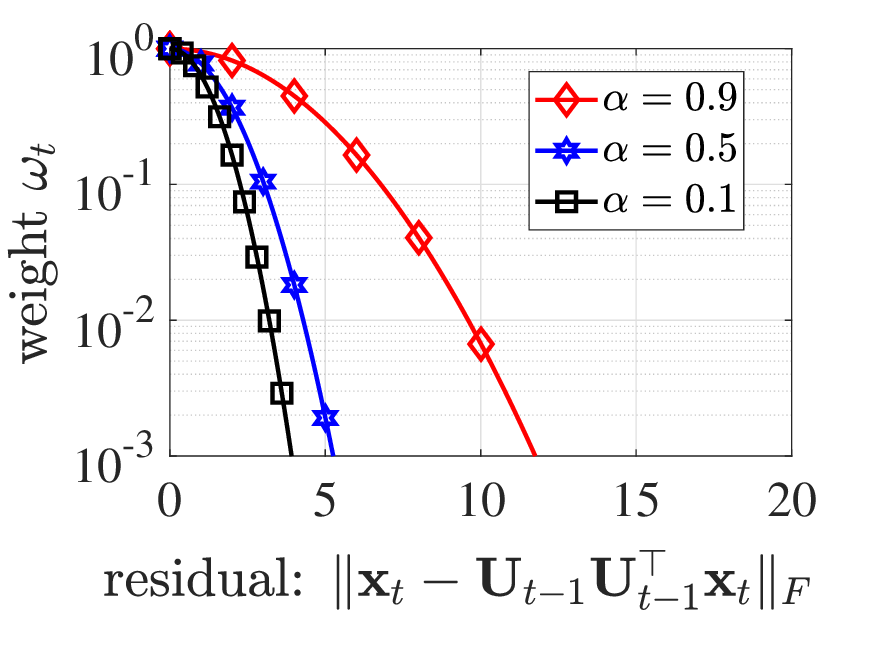}	\vspace{-0.5em}
		\caption{The weight $\omega_t$ is computed based on $\alpha$ divergence with $p = 2$.  When the residual is large, $\omega_t$ approaches zero; conversely, when the residual is small, $\omega_t$ approaches one.}
		\label{fig:alphadivergence}
		\vspace{2em}
	\end{figure}

	Following the robust statistical approach for sample covariance estimation proposed in~\cite{thanh2023aFAPI}, we first modify the ``scaled" covariance matrix~\eqref{eq:Rt}
	as follows
	\begin{align} \label{eq:robust_Rt}
		\mathbf{R}_t = (1-\lambda) \mathbf{R}_{t-1} + \lambda  \omega_t \mathbf{x}_t \mathbf{x}_t^\top,
	\end{align}  
	where $1-\lambda$ plays the same role as to the forgetting factor $\beta$ and
	the weight $\omega_t$ is chosen as 
	\begin{equation} \label{eq:weight}
	\omega_{t} = \exp\left( -\frac{1 - \alpha}{2} \big \| \mathbf{x}_{t} - \mathbf{U}_{t-1}\mathbf{U}_{t-1}^\top \x_t\big\|_F^p \right),
	\end{equation}
	with \( 0 < \alpha < 1 \) and \( 0 < p \leq 2 \). Here, $\U_{t-1}$ denotes the previous  estimate of the subspace matrix $\A$ at time $t-1$. The residual between the data observation and its projection is defined as $
		\| \e_t \|_F = \| \mathbf{x}_{t} - \mathbf{U}_{t-1}\U_{t-1}^\top \x_t \|_F.
	$ See Fig.~\ref{fig:alphadivergence} for an illustration.  A large residual suggests that $\x_t$ may not lie within the current subspace and could be a corrupted data sample. In such cases, the corresponding weight~$\omega_t$ becomes small (close to zero), reducing the influence of the corrupted data. 
	Conversely, as the residual approaches zero, $\omega_{t}$ approaches~$1$, indicating that
	the data sample is clean and should fully contribute to the subspace update. 
	
	The use of the weight $\omega_t$ in~\eqref{eq:weight} is motivated from the following observation. The signal model~\eqref{eq:data-model} implies the empirical distribution $g(\mathbf{x}, \mathbf{U})$  of data samples is a mixture of a true one $f(\mathbf{x}, \mathbf{U})$ (corresponding  to the low-rank signal) and a contaminated component, i.e., 
	\begin{align}
		g(\mathbf{x}, \mathbf{U}) = (1 - \delta) f(\mathbf{x}, \mathbf{U}) + \delta h(\mathbf{x}),
	\end{align}
	where \(h(\mathbf{x})\) represents impulsive or non-Gaussian noises and \mbox{$0<\delta<1$} is denotes a trade-off parameter between two distributions. Accordingly, the $\alpha$ divergence
	$
	\mathcal{D}_\alpha(g(\mathbf{x}, \mathbf{U}) \| f(\mathbf{x}, \mathbf{U}))
	$
	provides a robust estimation criterion  for estimating the underlying subspace as follows
	\begin{align} \label{eq:min-D_alpha}
	\A = \argmin_{\U}~\mathcal{D}_\alpha \Big\{ g(\mathbf{x}, \mathbf{U}\big) \big \| f(\mathbf{x}, \mathbf{U})\Big\}.
	\end{align} Since $g(\mathbf{x}, \mathbf{U}\big) $ is generally unknown in practice, the work \cite{rekavandi2023trpast} indicated that, in such cases, \eqref{eq:min-D_alpha} reduces to
	\begin{align}\label{eq:max-f-dist}
	\A & = \Scale[1]{\displaystyle \argmax_{\mathbf{U}} \frac{1}{1 - \alpha} \sum_{k = 1}^{t} f(\mathbf{x}_{k}, \mathbf{U})^{1 - \alpha}}. 
	\end{align} 
	 As indicated in our companion work~\cite{thanh2023aFAPI}, \eqref{eq:max-f-dist} is approximately equivalent to 
	 \begin{align} \label{eq:Min-weighted}
	 &	\A    = \argmin_{\mathbf{U}}  \sum_{k=1}^{t}    \widetilde{\omega}_k \left\| \mathbf{x}_k - \mathbf{U} \mathbf{U}^\top \mathbf{x}_k \right\|_F^2,   ~~~\text{with} \\
	 & 
	 \widetilde{\omega}_k  =  \exp\left( -\frac{1 - \alpha}{2} \big \| \mathbf{x}_{k} - \mathbf{U}_{k-1}\mathbf{U}_{k-1}^\top \mathbf{x}_{k} \big\|_F^2 \right).
	  \end{align} In parallel, minimizing a weighted least-square objective function with a weight $\widetilde{\omega}_k$ results in the principal subspace of the weighted covariance matrix $\C_k = \C_{k-1} + \widetilde{\omega}_k \x_k \x_k^\top$.  As a result, we can adopt the form~\eqref{eq:weight} to set the weight $\omega_{t}$ in \eqref{eq:robust_Rt}.  The inclusion of 
	  $p\leq 2$ . Accordingly, we reformulate the main step~\eqref{eq:OPIT-St} of the classical OPIT as follows
	  \begin{align}
	  	\S_{t}  =  (1- \lambda) \S_{t-1} \E_{t-1} +  \lambda \omega_t \x_{t} \w_{t}^\top.  
	  \end{align}
	  Other steps of $\alpha$OPIT can be given in Algorithm~\ref{al-aOPIT}.
	
	 \emph{Complexity Analysis}: $\alpha$OPIT shares the same computational complexity and memory storage as the classical OPIT method.  Specifically, its overall computational cost is $\mathcal{O}(nr^2)$ flops. The cost for each step of $\alpha$OPIT is provided in details in Algorithm~\ref{al-aOPIT}. In terms of memory storage, $\alpha$OPIT requires a total of $2nr + r^2$ words of memory to save $\U_t, \S_t$, and $\E_t$ at each iteration.  
	
	\section{Experiments}
	
	This section presents several experiments conducted to investigate the performance of $\alpha$OPIT. Its effectiveness is evaluated in comparison with several state-of-the-art  ST algorithms, including $\alpha$FAPI, TRPAST, ROBUSTA, and OPIT.  
	
	\subsection{Robust Sparse Subspace Tracking}
	
	\begin{figure}[!t]
		\centering
		\subfigure[Laplace noise]{\includegraphics[width=0.8\linewidth]{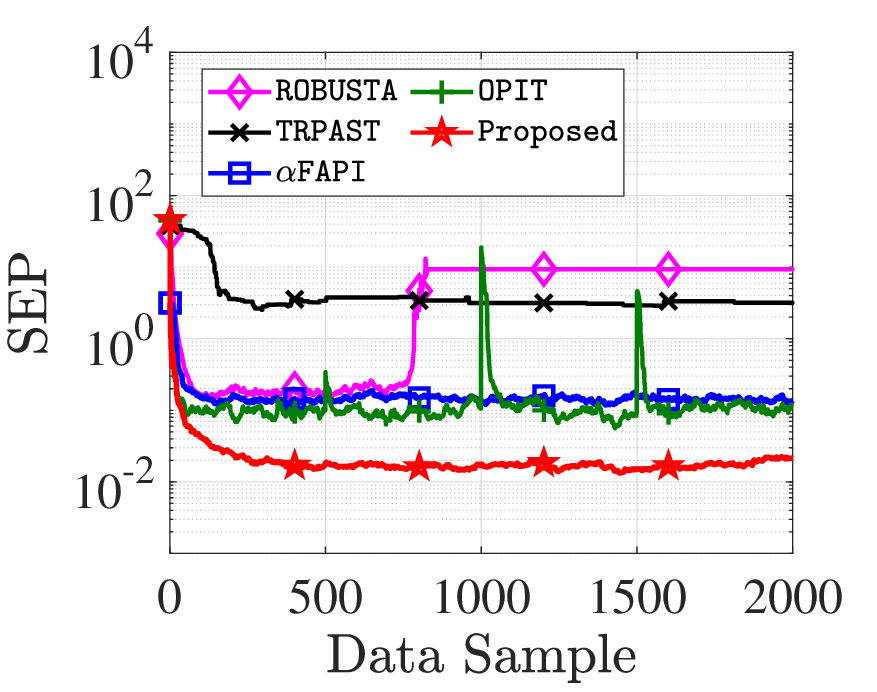}}\\[-0.5em]
		\subfigure[Cauchy noise]{\includegraphics[width=0.8\linewidth]{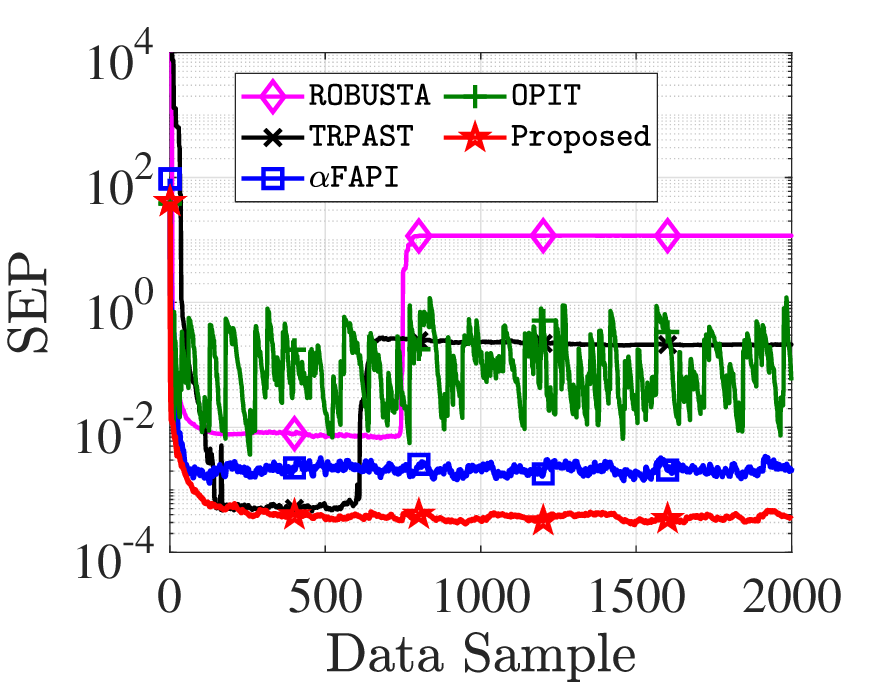}}\\[-0.5em]
		\subfigure[Laplace + Cauchy]{\includegraphics[width=0.8\linewidth]{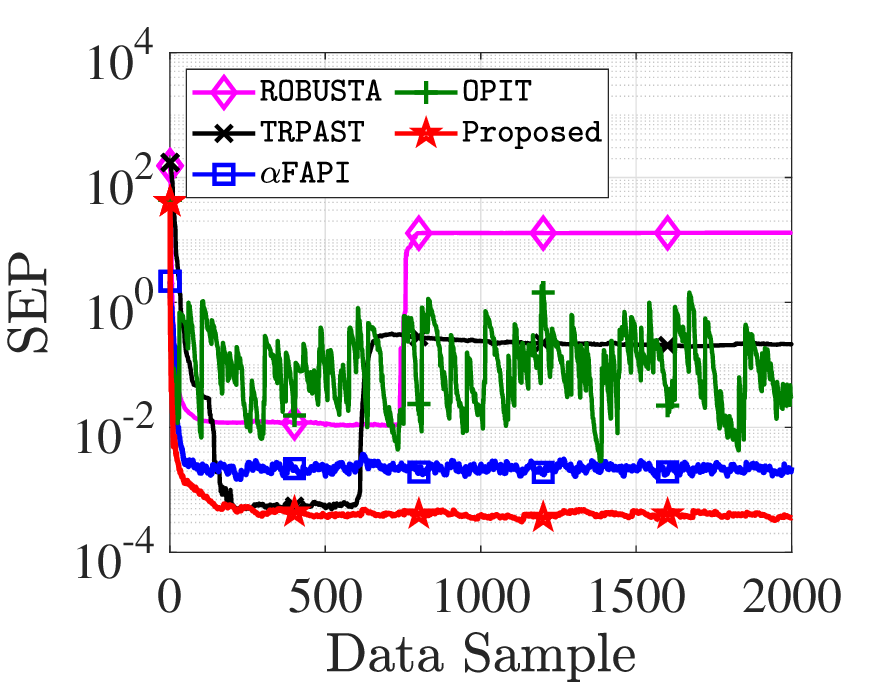}}
		
		\vspace{-0.75em}
		\caption{Experiment setup: data dimension $n=200$, true rank $ r= 5$, number of data samples $ T = 2000,$ time-varying factor $ \varepsilon_t = 10^{-2}$, subspace sparsity 80\%, the alpha divergence with~\mbox{$\alpha = 0.9$}.}
		\label{fig:rst}
		\vspace{1em}
	\end{figure}
	The signal samples \(\{\mathbf{x}_{t}\}_{t \geq 1}\) are generated based on the data model~\eqref{eq:data-model} where  \(\A =\mathbf{A}_{t}\) is a slowly time-varying sparse subspace matrix, generated recursively as follows
	\begin{equation}
	    \mathbf{A}_{t} = \bm \Omega \circledast (\mathbf{A}_{t-1} + \varepsilon_t \mathbf{V}_{t}),
	\end{equation}
	where $\bm \Omega \in \RR^{n\times r}$ is a binary matrix indicating the sparsity level of the subspace matrix, and  
	\(\mathbf{V}_{t} \in \RR^{n\times r}\) is Gaussian noise with zero mean and unit variance. The parameter \(\varepsilon_t\geq 0\) controls the level of subspace variation at each time step $t$. The vector \(\mathbf{w}_{t} \in \mathbb{R}^{r}\) denotes the coefficient vector. The vector \(\bm \nu_{t} \in \mathbb{R}^{n}\) accounts for non-Gaussian noises. 
	In our experiments, we consider the following three cases for the noise $\bm \nu_t$:
	\begin{align} 
	     \bm \nu_t(i) \sim &~ (1 - \delta) \mathcal{N}(0, \sigma_n^2) + \delta\, \mathrm{Laplace}(\mu, \gamma), \\
	    \bm \nu_t(i) \sim & ~(1 - \delta) \mathcal{N}(0, \sigma_n^2) + \delta\, \mathrm{Cauchy}(\mu, \gamma),  \label{eq:cauchy} 
	     \\
	      \bm \nu_t(i) \sim& ~ (1 - \delta) \mathcal{N}(0, \sigma_n^2) \notag \\
	      & +  \frac{\delta}{2}  \mathrm{Laplace}(\mu, \gamma)   +  \frac{\delta}{2}  \mathrm{Cauchy}(\mu, \gamma), 
	\end{align}
	where \(\delta\) represents the proportion of corrupted data, and their probability density functions are given by 
	\begin{align}
	   \mathcal{N}(x; \mu, \sigma_n^2)&= \frac{1}{\sqrt{2\pi \sigma_n^2}} \exp\bigg( -\frac{(x - \mu)^2}{2\sigma_n^2} \bigg), 
	  \\
	   \mathrm{Laplace}(x; \mu, \gamma) & = \frac{1}{2 \gamma} \exp\bigg( -\frac{|x - \mu|}{\gamma} \bigg), 
	 \\
	 \mathrm{Cauchy}(x; \mu, \gamma)    & = \frac{1}{ \displaystyle \pi \gamma \left[1 + \left( \frac{x - \mu}{\gamma} \right)^2 \right]}.  
	\end{align}
	Furthermore, two abrupt changes are made at $t=1000$ and $t=1500$ to evaluate the robustness of all ST algorithms. 
	
	To measure the accuracy of subspace tracking algorithms, we use the subspace estimation performance (SEP) metric, defined as follows.
	\begin{align}
	    \operatorname{SEP}\big(\U_{\text{true}},\U_{\text{est}}\big) = \frac{\operatorname{tr}\Big\{\U_{\text{est}}^\top \big( \I  - \U_{\text{true}} \U_{\text{true}}^\top \big)\U_{\text{est}}
	     \Big\} }{ \operatorname{tr}\Big\{\U_{\text{est}}^\top \big( \U_{\text{true}} \U_{\text{true}}^\top \big)\U_{\text{est}}
	     \Big\} }. \notag
	\end{align} The lower SEP indicates the better algorithm  performance. Fig.~\ref{fig:rst} illustrates the performance of all subspace tracking algorithms across three case studies involving non-Gaussian noise. As demonstrated, our method consistently outperforms other state-of-the-art robust subspace tracking methods in all settings. In particular, ROBUSTA and TRPAST are sensitive to both Laplace and Cauchy noise. Although OPIT performs well under Laplace noise, it exhibits performance degradation during abrupt changes at $t=1000$ and $t=1500$ and struggles with Cauchy noise.  $\alpha$FAPI can handle both types of corruption. However, its estimation accuracy is lower than that of our proposed method.

	\vspace{-0.5em}
	\subsection{Direction-of-arrival (DOA) Tracking}
	\vspace{-0.5em}

	We further evaluate the performance of the proposed $\alpha$OPIT algorithm in the context of direction-of-arrival (DOA) estimation for wireless communication systems. Assume that the received signal at each time 
	$t$ follows the data model
	\begin{align} \label{eq:DOA}
	\mathbf{x}_{t} = \mathbf{A}_{t}\mathbf{s}_{t} + \bm \nu_{t}.
	\end{align}
	Here, \(\mathbf{A}_{t} = [\mathbf{a}(\omega_{1,t}), \mathbf{a}(\omega_{2,t}), \ldots, \mathbf{a}(\omega_{K,t})]\)  is the time-varying steering matrix whose column is defined as \begin{align} \notag
		\a(\omega_{k,t}) = \big[1, \exp({j\omega_{k,t}}), \dots, \exp({j(n-1)\omega_{k,t}}) \big]^\top,
	\end{align} where \(\omega_{k,t} = \pi \sin \theta_{k,t}\) denotes the angular frequency associated with the DOA of the $k$-th user. The vector~\(\mathbf{s}_{t}\) represents the user signal, modeled as a complex Gaussian random vector with covariance matrix 
	 \(\mathbf{C}_s = \mathbf{I}_K\). The noise vector $\bm \nu_{t}$ follows the mixed distribution~\eqref{eq:cauchy}  described in the previous task of robust subspace tracking. 
	 
		\begin{figure}[!t]
		\centering
		\includegraphics[width=0.85\linewidth]{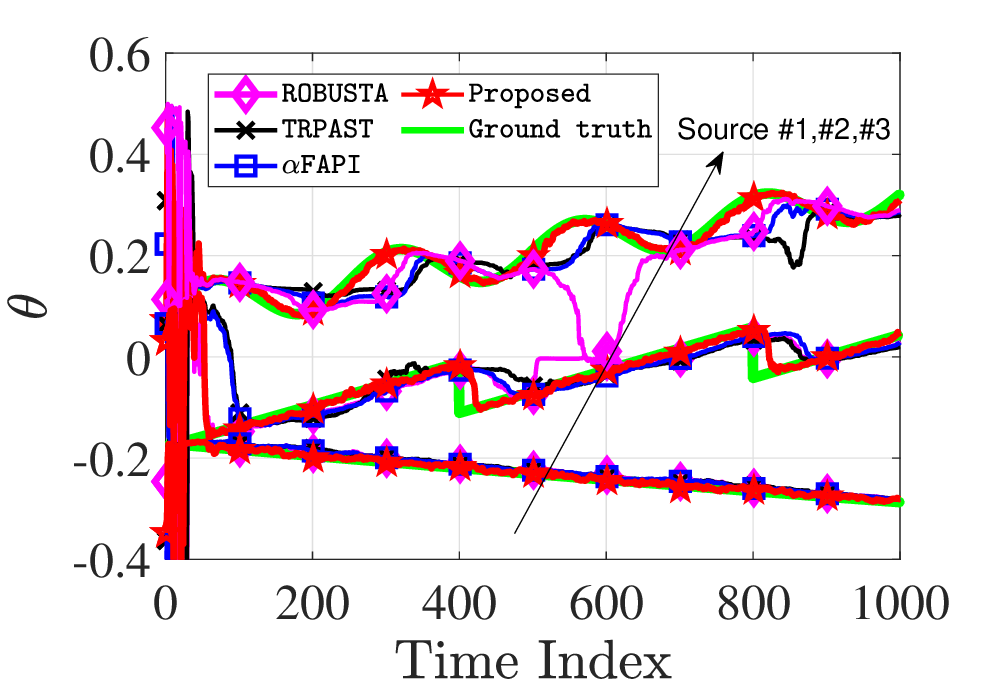}
		\vspace{-0.5em}
		\caption{DOA tracking  setup:  $n=20$ sensors,  $r=3$ signal sources, and  $T=1000$ data samples. Alpha divergence with $\alpha = 0.9$ is used. The three underlying DOAs consist of a linear source ($\#1$), a sawtooth source ($\#2$) and a sinusoidal source ($\#3$), illustrated by (\protect \hwplotD).}
		\label{fig:DOA}
		\vspace{1em}
	\end{figure}

	The main objective is to estimate the DOAs from the corrupted observations \(\mathbf{x}_t\). To this end, $\alpha$OPIT is applied to recursively and robustly estimate the underlying signal subspace.  Once the subspace is obtained, the ESPRIT algorithm~\cite{32276} is applied to extract angular frequencies \(\omega_k(t)\), and thus recover the DOAs \(\theta_k(t)\).

	The experimental results in Fig.~\ref{fig:DOA} demonstrate that $\alpha$OPIT achieves high accuracy in DOA estimation under  non-Gaussian noise conditions. Compared to existing algorithms such as ROBUSTA, TRPAST, and $\alpha$FAPI, our method consistently offers superior robustness for the DOA tracking task. Notably, in scenarios where the angular frequency $\theta$ changes fast (e.g., source $\#1$ or source $\#2$ at $t=400$ and $t=800$),  $\alpha$OPIT  adapts more quickly and tracks the variations more accurately than the competing methods. 

	

    \section{Conclusions}	 
	In this paper, we addressed the problem of robust subspace tracking with data corruption. We proposed a novel robust and adaptive algorithm, called $\alpha$OPIT, for tracking the principal sparse subspace of streaming data over time. Experimental results demonstrated that $\alpha$OPIT effectively handles various types of data contamination and consistently outperforms existing   subspace tracking algorithms. Future works will explore its performance in high-dimensional settings and fast time-varying cases. In addition, when handling higher-order streaming data, subspace tracking naturally extends to tensor tracking~\cite{thanh2023contemporary}. Therefore, exploring robust tensor tracking through $\alpha$-divergence is a promising direction for future research.  
	

	 {

	 \bibliographystyle{ieeetr}
	 \small 
	 	\balance 
	 \bibliography{references}
	}

	\end{document}